\documentclass[fleqn,usenatbib]{mnras}

\usepackage{newtxtext,newtxmath,booktabs}

\usepackage[T1]{fontenc}

\DeclareRobustCommand{\VAN}[3]{#2}
\let\VANthebibliography\thebibliography
\def\thebibliography{\DeclareRobustCommand{\VAN}[3]{##3}\VANthebibliography}

\usepackage{graphicx}	
\usepackage{amsmath}	

\title[Biological evolution on Hycean worlds]{Prospects for biological evolution on Hycean worlds}

\author[E. G. Mitchell and N. Madhusudhan]{
Emily G. Mitchell,$^{1,2}$\thanks{E-mail: ek338@cam.ac.uk}
Nikku Madhusudhan,$^{3}$\thanks{E-mail: nmadhu@ast.cam.ac.uk}
\\
$^{1}$Department of Zoology, University of Cambridge, Downing Street, Cambridge, CB2 3EJ, UK\\
$^{2}$University Museum of Zoology Cambridge, Downing Street, Cambridge, CB2 3EJ, UK\\
$^{3}$Institute of Astronomy, University of Cambridge, Madingley Road, Cambridge CB3 0HA, UK.
}

\date{Accepted XXX. Received YYY; in original form ZZZ}

\pubyear{\the\year{}}

\begin{document}
\label{firstpage}
\pagerange{\pageref{firstpage}--\pageref{lastpage}}
\maketitle

\begin{abstract}
Recent detections of carbon-bearing molecules in the atmosphere of a candidate Hycean world, K2-18 b, with JWST are opening the prospects for characterising potential biospheres on temperate exoplanets. Hycean worlds are a recently theorised class of habitable exoplanets with ocean covered surfaces and hydrogen-rich atmospheres. Hycean planets are thought to be conducive for hosting microbial life under conditions similar to those in the Earth's oceans. In the present work we investigate the potential for biological evolution on Hycean worlds and their dependence on the thermodynamic conditions. We find that a large range of evolutionary rates and origination times are possible for unicellular life in oceanic environments for a relatively marginal range in environmental conditions. For example, a relatively small (10 K) increase in the average ocean temperature can lead to over twice the evolutionary rates, with key unicellular groups originating as early as $\sim$1.3 billion years from origin of life. On the contrary, similar decreases in temperatures can also significantly delay the origination times by several billion years. This delay in turn could affect their observable biomarkers such as dimethylsulfide, which is known to be produced predominantly by Eukaryotic marine phytoplankton in Earth's oceans. Therefore, Hycean worlds that are significantly cooler than Earth may be expected to host simpler microbial life than Earth's oceans and may show weaker biosignatures, unless they orbit significantly older stars than the Sun. Conversely, Hycean worlds with warmer surface temperatures than Earth are more likely to show stronger atmospheric biosignatures due to microbial life if present. 

\end{abstract}




\section{Introduction}
The search for extraterrestrial life is one of the most fundamental quests in human history. It has the potential to both ascertain the uniqueness or ubiquity of terrestrial-like life in the universe as well as provide insights into the conditions that led to the origin of life here on Earth. Current advances in astronomical observations indicate a realistic possibility for detecting life beyond the solar system in the foreseeable future. While an exact Earth-twin around a sun-like star is yet to be discovered, a number of temperate Earth-size and sub-Neptune exoplanets have been discovered around M dwarf host stars smaller and cooler than the sun and, hence, with habitable zones that are significantly closer to the stars \citep[e.g.][]{montet2015,gillon2017,dittmann2017}. Furthermore, the advent of new observational facilities, such as the James Webb Space Telescope (JWST), and the upcoming extremely large telescopes on ground promise the capability for detection of atmospheric signatures, including biomarkers, in such planets \citep{barstow2016,wunderlich2019,madhusudhan_habitability_2021}. 

\begin{figure}
         \includegraphics[width=0.5\textwidth]{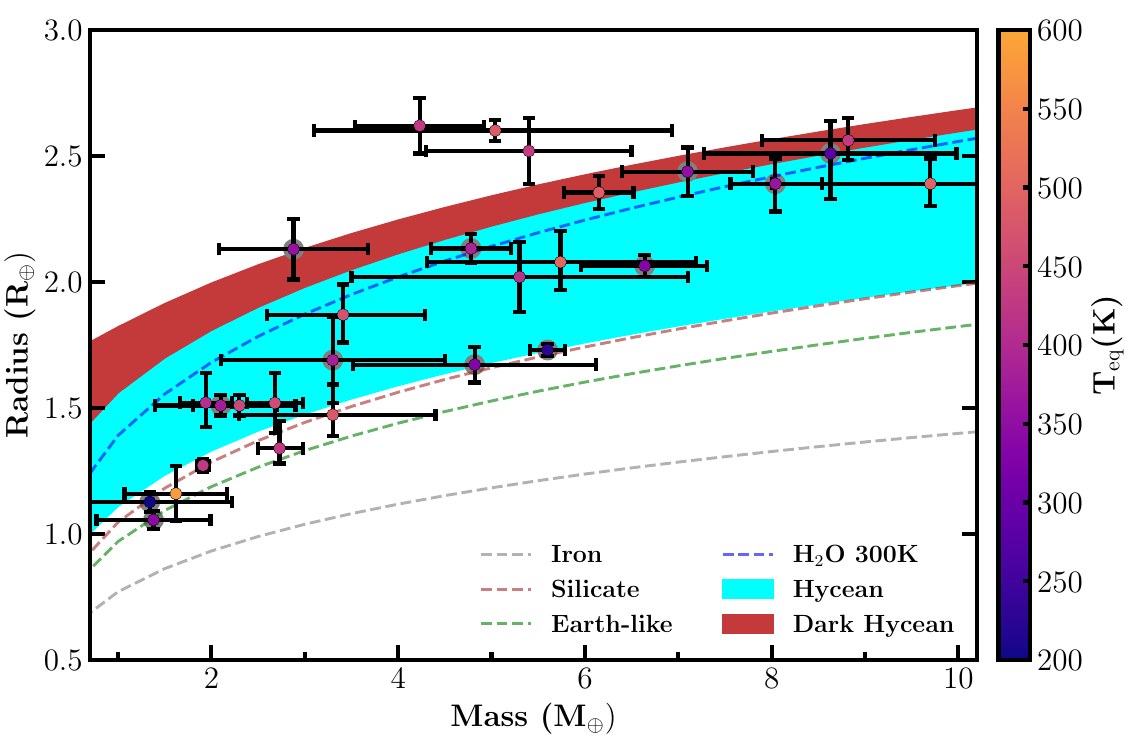}
         \caption{Bulk properties of temperate sub-Neptune exoplanets. The circles with uncertainties show a selected sample of temperate sub-Neptune exoplanets with confirmed measurements of radii and masses, with uncertainties below 2 earth masses, and zero-albedo equilibrium temperatures (T$_{\rm eq}$) below 600 K. The circles are color-coded by T$_{\rm eq}$ as denoted by the color bar. The dashed lines show theoretical mass-radius curves of model planets of uniform interior compositions denoted in the legend  and the cyan and crimson regions denote ranges of masses and radii allowed for Hycean and Dark Hycean planets, following \citet{madhusudhan_habitability_2021}. A subset of these planets with T$_{\rm eq}$ below $\sim$400 K and lying in the cyan region, marked with grey outer circles, could be candidate Hycean planets \citep{madhusudhan_habitability_2021}. Adapted from \citet{madhusudhan_habitability_2021} with data updated based on the NASA Exoplanet Archive.} 
         \label{fig:mr}
\end{figure}

An important recent development in this direction is the possibility of Hycean worlds, which increase both the numbers of potentially habitable planets and the ability to detect biosignatures in their atmospheres \citep{madhusudhan_habitability_2021}. Hycean worlds are a recently proposed class of planets with ocean covered surfaces and hydrogen-rich atmospheres. Their volatile-rich interiors lead to larger sizes and extended atmospheres, compared to rocky planets of similar mass, and make them more conducive to atmospheric observations. Several candidate Hycean planets have been identified among the currently known temperate sub-Neptunes, as shown in Fig.~\ref{fig:mr}, that are good targets for atmospheric observations with JWST \citep{madhusudhan_habitability_2021}. A major observational breakthrough came with the recent detection of carbon bearing molecules, methane and carbon dioxide, in the candidate Hycean world K2-18b using JWST \citep{madhusudhan_carbon-bearing_2023}. Such observations and theoretical studies demonstrate the capability for detection of biomarkers, if life is indeed present on such planets, using current observational facilities in the near future. 

Hycean planets are expected to provide the chemical and thermodynamic conditions necessary for sustaining oceanic microbial life \citep{madhusudhan_bioessential_2023, Glein2024, Petraccone2024}. The H$_2$-rich atmospheres provide a rich source of organic prebiotic molecules in the early evolution of such planets while the bioessential elemental requirements, including CHNOPS elements, can be met through a combination of external delivery and atmospheric precipitation \citep{madhusudhan_bioessential_2023}. The availability of abundant H$_2$ and oxidised compounds such as CO$_2$ can release significant metabolic energy through reductive reactions as well as the synthesis of complex hydrocarbons, including amino acids \citep{Glein2024}. At the same time, the wide range of temperatures possible on such planets allow for diverse thermodynamic conditions to support life. In particular, most of the Hycean candidates currently known are expected to be warmer than average terrestrial conditions, leading to higher entropy and more conducive conditions for life \citep{Petraccone2024} than Earth-like rocky exoplanets in the habitable zone. 

In the present study we investigate the prospects for biological evolution on Hycean worlds. Previous studies have shown that increased temperatures in habitable environments are generally more conducive to biological activity \citep{Petraccone2024} and lead to increased evolutionary rates and higher diversity \citep{lingam2018}. We investigate the degree to which changes in the ocean surface temperature would affect the evolution of key groups of Earth-like unicellular life and their origination timescales in a Hycean environment. This work, in turn, has observable consequences for prominent biosignatures on such planets, considering that unicellular phytoplankton are a major source of key biomarkers in the Earth's atmosphere, such as dimethyl sulphide which may be observable in Hycean atmospheres \citep{madhusudhan_habitability_2021}. 

\section{Modelling biological evolution on Hycean planets}
We assess the nature of unicellular life one may expect, i.e. what makes up the composition of the biosphere, if life originated on a Hycean planet assuming Earth-like organisms, i.e. bacterial, archaea and eukaryotic life. Biosphere composition depends on what organisms have evolved at a given point in time. For example there were no animals 4 billion years ago whereas today they dominate the biosphere. As such, the key aspect we wish to understand for a given Hycean exoplanet is what unicellular organisms could have evolved by a given time, i.e. what are the origination times for different major groups. However, the drivers that shape biological macroevolution are not always well resolved, especially for the origination of new groups, e.g. the origins of animals \citep{hoffman_snowball_2017,cavalier2017origin,sogabe2019pluripotency,liu2024marine,stockey2024sustained}. This uncertainty behind evolutionary drivers makes predicting the evolutionary histories of life on exoplanets hard since there are not many direct causal relationships between biosphere conditions and origination of important clades (the group of species that includes the last common ancestor and all its living and extinct descendants). Yet often macroevolutionary patterns can be well modelled through neutral models of evolution, i.e. without incorporating causal biotic or environmental drivers into the models.  Such models can capture large-scale patterns because without any environmental or biotic pressures, populations still change, and species evolve and go extinct through processes such as neutral molecular evolution \citep{kimura1989neutral,ohta1992nearly}. 

Neutral evolution models provide good empirical match to observed data on Earth across different temporal scales. On planetary time scales, simple birth-death models of macroevolutionary patterns, whereby species evolve and go extinct with given probabilities, replicate the broad diversification patterns in large groups such as arthropods, tetrapod and land plants \citep[e.g.][]{budd_dynamics_2020}. Similarly, early animals appear to have neutral dynamics, in that they only have limited biotic interactions and interactions with their local environment \citep{mitchell_importance_2019}.    As such, the simplest models are these neutral models which not only give a good indicator of speciation rates without assuming any causal drivers, but also likely are the most conservative, i.e. likely provide the slowest rate of evolution and speciation rates.

Evolutionary rates can be measured (and modelled) in different ways, including speciation rates through geological time or the number of nucleotide substitutions due to random mutations i.e. neutral molecular evolution. One approach to modelling evolutionary rates is using the metabolic theory of ecology (MTE) to predict the evolutionary rates based on the metabolic rate of an organism \citep{allen2006kinetic}. The MTE provides a framework for predicting biological traits, particularly metabolic rate, for a given organism \citep{brown2004toward}. These predictions match empirical observations well \citep{brown2004toward}, albeit the underlying mechanisms for this relationship are debated (e.g. \citet{maino2014reconciling,west1997general,price2012testing,o2007reconsidering}).  The MTE builds on metabolic scaling theory, which states that the metabolism of organisms is \textit{the} fundamental biological rate which drives most observed patterns within biology and depends on the body size of the organism and the temperature of its environment \citep{west1997general,brown2004toward,gillooly2001effects}. Body-size changes metabolic rates due to surface-volume ratios because smaller organisms need more energy (per unit biomass) to transfer nutrients around their body than larger organisms.  Metabolism is impacted by temperature because temperature will change the rate at which all biochemical/physiological reactions occur within an organism, thus speeding up rates of biochemical reactions. The MTE links the organism-level traits of metabolic scaling theory and metabolic rates to life-history and ecological traits such as population growth, carrying capacity and biodiversity \citep{brown2004toward,mccoy2008predicting} including large scale patterns in current global diversity patterns such as the latitudinal diversity gradients  \citep{righetti2019global}.

The key variables that predict the temperature-corrected metabolic rate for an organism are the environmental temperature, the mass of the organism and the mass scaling component  \citep{brown2004toward}, as shown in Eq \ref{equ:mte} in section~\ref{sec:methods}. The mass scaling component is predicted by MTE and the Metabolic scaling theory as ${-\frac{1}{4}}$ for the temperature corrected metabolic rate, and has been repeatedly validated against empirical data \citep[e.g.][]{raven1988temperature, vetter1995ecophysiological,mcclain2012energetics}.  There is a substantial amount of empirical evidence for the relationship between body-size, temperature and metabolic rates, consistently finding these relationships across all kingdoms of life, in different environments and across 50 orders of magnitude of body-size, from bacteria to large whales \citep{brown2004toward,munch2009latitudinal,gillooly2017broad}. The largest deviations tend to include endothermic organisms, likely because they do not take on the ambient temperature of their environment such that birds, mammals and some fish have greater deviations from the MTE predictions \citep{brown2004toward,grady2014evidence}.  Most of this work testing the MTE has focused on the 0-40$^o$C range, but there is limited data for extremophiles such as hydrothermal vent organisms and deep-sea organisms \citep{mcclain2012energetics}, which also broadly fit the data. Empirical support for the life-history, population and ecosystem level traits predicted by  \citet{brown2004toward} are more variable, so for this study we focus solely on evolutionary rates. 

In terms of the use of MTE to make predictions about evolution of life on exoplanets, \citet{lingam2018} noted that because of the relationships predicted by MTE, once life has originated, the chance of complex life is greater on warmer planets because of increased evolutionary and speciation rates and also because of the greater biodiversity that comes with higher temperatures. In this study, we use the MTE to make quantitative inferences of the extent to which evolutionary rates change with temperature, and use these rates to infer how the time to the origination of clades will also change. 

\section{Methods}
\label{sec:methods}
The evolution rate of an organism is inversely proportional to the time to speciation and is proportional to metabolic rate \citep{allen2006kinetic}. The relationship of metabolic rate with evolutionary and speciation rates have been verified using the model group foraminifera using speciation rates derived from fossil and extant foraminifera \citep{allen2006kinetic}. Therefore, by calculating the predicted metabolic rate of an organism, the evolutionary rates and time to speciation can also be inferred. 

In order to use the MTE to estimate how much time is required on different planets for the origination of major groups to evolve, we first need to estimate how much ``evolution" was required here on Earth.  The amount of required evolution is taken to be (for given sized organism) how many nucleotide substitutions (mutations) are needed here on Earth to go from the last common ancestor of all organisms to the origination of a new clade. These evolutionary relationships can be encapsulated by an evolutionary tree, or time-calibrated phylogeny where the patterns of the branches of the tree reflects the relationships between different species (or groups of species) and where the lengths of the branches depict time. 

To generate these evolutionary trees for Hycean conditions, we make some minimal assumptions (see Appendix~\ref{appendix:assumptions} for more details): 1) Life has originated (Origin of Life, OoL).  2) Elements/resources are not limiting factors, as suggested in \citet{madhusudhan_habitability_2021}, so that the biosphere is stable and life is maintained. 3) Speciation leads to the origination of new clades similar to that in Earth's oceans.  4) The planetary temperature isn't significantly changed by the organisms considered. While the extent to which a Hycean world would satisfy these conditions is not yet known, these nevertheless provide a starting point to assess the prospects for evolution of Earth-like unicellular life on such planets. 

Evolutionary rates at a given point in time are calculated based on the temperature-corrected metabolic rate for the given organism: 
\begin{equation}\label{equ:mte}
    \bar B =b_0 M^\beta e^{\frac{-E}{kT}}
\end{equation}

Where $\bar B $ is the temperature-corrected metabolic rate, $\beta$ is the mass scaling component, $b_0$ is the  taxa dependant constant, and for unicellular organisms $ln(b_0)= 15.85 $ \citep{brown2004toward}. The activation energy $E$ can vary between 0.2 and 1.2 eV, but consistently has an observed value in this MTE context of 0.65 eV \citep{raven1988temperature, vetter1995ecophysiological}, 1 eV = 1.602 $\times 10^{-19}$ J. Here, $k$ = $8.62 \times 10^{-5}$ eV K$^{-1}$ is the Boltzmann constant, $M$ is mass in grams and $T$ is the absolute temperature in K. The mass scaling component $\beta$ is predicted by MTE and the Metabolic scaling theory as -1/4.  

In this study the input parameters in Eq (\ref{equ:mte}) were complied from \citep{brown2004toward} using values for unicellular organisms.  The input masses for the different species are calculated from the volume, given in Table \ref{tab:masses} which gives the median volume for the given organism as recorded in the literature. The masses were calculated as the volume times the density of phytoplankton of 1.1 g cm$^{-3}$ \citep{padisak2003sinking}.

\begin{table*}
    \centering
    \caption{Summary of the organisms and their properties considered for calculations of evolutionary rates. Node origination times are given by the median values of published works with the range of uncertainties given in parentheses.}
    \begin{tabular}{llcccc}
\textbf{Clade} & \textbf{Example Organism}  & \textbf{Size (m$^3$)}& \textbf{Reference} & \parbox[t]{3cm}{\textbf{Node origination time}\\ \textbf{from Present (Myr)} } & \textbf{Reference} \\
Cyanobacteria & \textit{Prochlorococcus} & 9.06$\times 10^{-19}$ & \citep{partensky1999prochlorococcus} & 2810 (2561 - 3306) & \citep{kumar2022timetree}\\
Aquificota & \textit{Aquifex} & 1.57$\times 10^{-18 }$& \citep{guiral2012hyperthermophilic} & 4101 (3644 - 4189) & \citep{kumar2022timetree}\\
Alphaproterobacteria & \textit{Alphaproterobacteria} & 1.77$\times 10^{-18 }$& \citep{brown2012polar} & 1900 (2423 – 1419) & \citep{wang2021dating}\\
Betaproterobacteria & \textit{Betaproterobacteria} & 4.91$\times 10^{-20}$ & \citep{croue2013single} & 1900 (2423 – 1419) & \citep{wang2021dating}\\
Gammaproteobacteria & \textit{Thiomargarita} & 3.35$\times 10^{-11}$ & \citep{volland2022centimeter} & 1900 (2423 – 1419) & \citep{wang2021dating}\\
Thermoplasmatales & \textit{Thermoplasmatales} & 4.19$\times 10^{-18}$ & \citep{paul2012methanoplasmatales} & 1704 (1132 - 2276) & \citep{kumar2022timetree}\\
Thermococcus & \textit{Thermococcaceae} & 1.72$\times 10^{-17}$ & \citep{canganella1998thermococcus} & 1576 (974 - 2410) & \citep{kumar2022timetree}\\
Methanococccea & \textit{Methanococcaceae} & 3.53$\times 10^{-18}$ & \citep{jones1983methanococcus} & 1576 (974 - 2410) & \citep{kumar2022timetree}\\
Archaeoglobus & \textit{Archaeoglobaceae} & 5.24$\times 10^{-19}$ & \citep{slobodkina2021physiological} & 1576 (974 - 2410) & \citep{kumar2022timetree}\\
Opisthokonta & \textit{Choanoflagellate} & 8.71$\times 10^{-17 }$& \citep{mah2014choanoflagellate} & 1598 (1085 - 1671) & \citep{kumar2022timetree}\\
Excavata & \textit{Jakoba  libera} & 3.14$\times 10^{-16}$ & \citep{patterson1993heterotrophic} & 1598 (1085 - 1671) & \citep{kumar2022timetree}\\
Alveolata & \textit{ciliates} & 1.13$\times 10^{-13}$ & \citep{nielsen1994regulation} & 1040 (627 - 1344) & \citep{kumar2022timetree}\\
Stramenopiles & \textit{Sagenista} & 1.13$\times 10^{-16}$ & \citep{cho2024phylogenomic} & 1106 (954 - 1234) & \citep{kumar2022timetree}\\
Rhizaria & \textit{Polycystinea} & 1.77$\times 10^{-12}$ & \citep{smalley1963radiolarians} & 933 (812.4 - 1470) & \citep{kumar2022timetree}\\
Diatoms & \textit{Chaetoceros} & 2.68$\times 10^{-18}$ & \citep{tomas1997identifying} & 206 (range unavailable) & \citep{kumar2022timetree}\\
Coccolithophores & \textit{Emiliania huxleyi} & 1.13$\times 10^{-16}$ & \citep{hoffmann2015insight} & 280  (range unavailable)& \citep{kumar2022timetree}\\
Dinoflagellates & \textit{Pfiesteria piscicida} & 5.96$\times 10^{-12}$ & \citep{burkholder1997pfiesteria} & 239  (range unavailable)& \citep{kumar2022timetree}\\
    \bottomrule
    \end{tabular}
    \label{tab:masses}
\end{table*}

\citet{allen2006kinetic} derived how the evolutionary rates follow from Eq (\ref{equ:mte}), showing the mutation rate as the number of nucleotide substitutions per unit time (cf. neutral molecular evolution) 
\begin{equation}\label{equ:mu}
    \alpha = \alpha_0 \bar B,
\end{equation}

where the mutation rate $\alpha$ can vary depending on clade for given incident radiation, but can be assumed to be constant for a given clade. This equation can then be used to derive the time it takes for speciation to occur, $t_s$:
\begin{equation}\label{equ:ts}
    t_s \propto 1/B_o e^{\frac{-E}{kT}}
\end{equation}

The origination times of different clades on Earth are dependent on the body-size of the organism, planetary temperature and the amount of neutral molecular evolution (the amount of background mutations). For the Earth's temperature, we use the median temperature values as calculated by \citet{krissansen2018constraining} at 0.01 Gyr intervals from the Origins of Life (OoL) at 4.3 billion years ago \citep{kumar2022timetree} to the present.   Required inputs for Eq (\ref{equ:mte}) are the body-masses of the organisms \ref{tab:masses}, with the other parameters adopted from the literature \citep{brown2004toward}.  Origination times are given using Time Tree \citep{kumar2022timetree} which is a meta analysis of published molecular clock studies which provides the origination times as the median values of published works (Table \ref{tab:masses}). Alphaproterobacteria, Betaproterobacteria and Gammaproterobacteria origination times were not given by \citet{kumar2022timetree} so we used \citet{wang2021dating} for these three clades. We used Kumar et al. 2022 rather than multiple sources so as to enable a consistent treatment for as many clades as possible. The timings along with the calculated evolutionary rates were input into the phylogeny and plotted with the R package ggtree \citep{yu2020using}.

In order to calculate the origination time of a group, the cumulative amount of substitutions needed for the origination of a group is calculated as the integral of Equation \ref{equ:mte} along all the ancestral branches. To calculate how the origination times for the clades change for different median temperatures (0, $\pm$ 5 K ,$\pm$ 10 K) the total number of mutations needed $\alpha_T$ was calculated as the sums of the integrals of Eq. \ref{equ:mu} for each ancestral branch of the phylogeny up to the origination times of the clade. At each new temperature, the number of mutations occurring up to each ancestral node was calculated in sequence up until the point that $\alpha_T$ was reached, and the corresponding time represented the origination time. 

In this study, we model the evolutionary rates and origination times for the major clades across the tree of life, namely Terrabacteria and Pseudomondoata clades within Bacteria, Euryarchaea and Thermoplasmatales within Archaea and SAR, Opisthokonta and Excavata with the Eukaryotes.  Within each of these clade we use as a representative organism a unicellular, holoplanktonic species (organisms which live their entire life-cycle in the water column) (Table \ref{tab:masses}).  While our analyses are focused on the normal thermal tolerance range of 0-40$^{o}$ C because the majority of life lives within this range \citep{sunday2012thermal}, the clades in our analyses include extremophiles, within Archaea, especially Euryarchaea containing numerous hyperthermophiles ($>60^o$C) and bacteria with thermophiles ($>40^o$C) \citep{dalmaso2015marine}.

Other groups of interest include ones likely to produce biosignatures, such as dimethylsulfide (DMS). DMS is produced biogenically, predominantly by marine phytoplankton as a breakdown of dimethylsulfoniopropionate (DMSP) which is an osmoregulator for phytoplankton \citep{andreae1990ocean,yoch2002dimethylsulfoniopropionate}.  DMS (and DMSP) are released by phytoplankton through growth, as an excretion product and when they die (or are eaten) DMSP is released and converted to DMS \citep{kwint1995dimethylsulphide}.  DMS can be produced by a wide range of phytoplankton including bacteria such as Cyanobacteria and Gammaproterobacteria \citep{zheng2020bacteria,teng2021biogeographic} but the majority of DMS in today's oceans are produced by dinoflagellates, diatoms, coccolithophores and protists \citep{hopkins2023biogeochemistry,yoch2002dimethylsulfoniopropionate}, i.e. Eukaryotes.  In order to further explore the diversity of phytoplankton (photosynthetic autotrophs) on Hycean worlds we also model the origination times of the major phytoplankton groups, Cyanobacteria, Gammaproterobacteria (within the group Pseudomonadota), Dinoflagellates, Coccolithophores and Diatoms (Table \ref{tab:masses}). Animals such as corals can also produce DMS through their symbionts \citep{jackson2020dmscorals}, but due to their benthic nature are beyond the scope of this paper.

In order to assess how our results may change using the recent origination times of \citet{moody2024nature}, we used their origination times for the last universal ancestor for all (4.2 Ga) and common ancestors for Eukaryotes (2.3 Ga), Archaea (3.5 Ga), Bacteria (3.4 Ga) and oxyphotobacteria (3.0 Ga) using the divergence times for the ILN, concatenated and cross bracing A model. 

\section{Results}
We used the MTE model to assess the effect of temperature on the evolutionary rate and origination times of key unicellular and phytoplankton groups discussed above. To illustrate the impact of median surface temperature on evolutionary rates, we calculated the evolutionary rates for an example methanogen over 4.3 billion years for surface temperature changes of $\pm$5 K, $\pm$10 K and $\pm$15 K, relative to Earth. We then calculated the origination times for the key unicellular  and phytoplankton groups for surface temperature changes of $\pm$ 10 K relative to Earth. 
\subsection{Evolutionary rates}
In order to illustrate how evolutionary rates change with temperature over planetary timescales we have calculated the evolutionary rates for an example organism (\textit{Aquifix}) over the last 4.3 billion years. \textit{Aquifix} is a suitable model organism because it is a good analogue to some of the first suggested life on Earth \citep{dodd2017evidence} in that it has similar ecological and morphological properties such as size. In order to aid comparison of the relative speed of the evolutionary rates we have normalised the rates so that the evolutionary rate at the Earth's temperature OoL 4.3 billion years ago (i.e. at \textit{t} = 0) is set to one. 

\begin{figure}
        \includegraphics[width=0.48\textwidth]{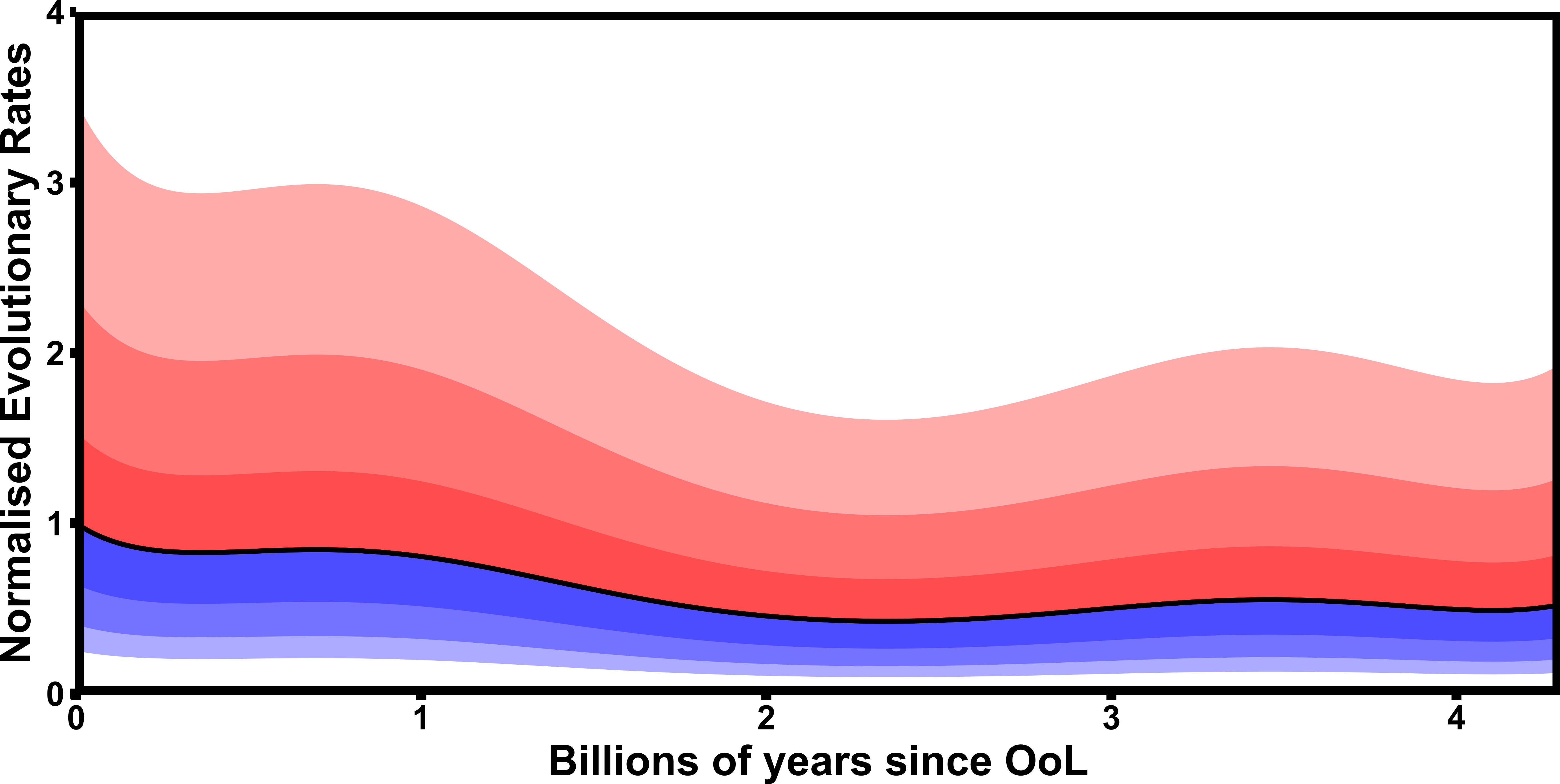}
        \caption{Effect of temperature on normalised evolutionary rate for an analogue LUCA methanogen on Earth. Cases with increased temperatures by +5 K, +10 K and +15 K, relative to Earth, are shown in different shades of red, and cases with decreased temperatures of -5 K, -10 K and -15 K, relative to Earth, are shown in blue. \textit{t} = 0 is set to when life is inferred to have originated, and the evolutionary rates normalised such that evolutionary rate is 1  at \textit{t} = 0 for Earth.}
        \label{fig:EarthTemps}
\end{figure}

Figure \ref{fig:EarthTemps} shows how the normalised evolutionary rate changes through time, with the fluctuations corresponding to the recorded median surface temperature through Earth’s history \citep{krissansen2018constraining}. The evolutionary rate decreases through time as the median surface temperature of the Earth decreases from 294 K at \textit{t} = 0 to 286 K for the present time, with a present evolutionary rate of 52.2\% of the starting rate, i.e. that at \textit{t} = 0. The mean normalised evolutionary rate over the last 4.3 Gyr is 60\% of the starting rate. When the median surface temperature of a planet increases, the evolutionary rate also increases. An increase of 5 K results in a faster evolutionary rate of 153\% at \textit{t} = 0, higher present rates of 82\% and a higher mean of 94\% over 4.3 Ga. An increase of 10 K corresponds to a faster evolutionary rate of 232\% at \textit{t} = 0, 127\% in the present and a mean increase of 145\% over 4.3 Ga.  An increase of 15 K corresponds to a faster evolutionary rate of 348\% at \textit{t} = 0, 193\% in the present and a mean increase of 220\% over 4.3 Ga.  Decreasing the median temperature decreases the evolutionary rates, with a 5 K decrease leading to rates of 64\% at \textit{t} = 0, and 32\% in the present, with a mean of 38\%. A decrease of median temperature by 10 K further reduces rates to 40\% at \textit{t} = 0, and 20\% in the present, with a mean of 24\% and a decrease of 15 K to 25\% at \textit{t} = 0, and 12\% in the present, with a mean of 14\%.  

The relationship between surface temperature changes and evolutionary rates are non-linear, following an exponential relation as seen from Eq (\ref{equ:mte}). This non-linearity results in much larger and more variable impacts of high temperatures on evolutionary rates than for colder temperatures. 

The range of surface temperatures that Fig. \ref{fig:EarthTemps} covers is from 309 K (36$^{o}$ C) for the +15 K case at \textit{t} = 0 to the coldest of 271K (-2$^{o}$ C) at -15 K.  The only scenario when the temperature is subzero is for the last 1.52 Gyr for the -15 K model run.  The thermal tolerance for life without extremophile adaptations is 0-40$^{o}$ C \citep{sunday2012thermal}, thus it is only for the coldest run that caution is needed when considering canonical life.

\subsection{Origination times of major groups}
\begin{figure*}
       \centering
         \includegraphics[width=17.8cm]{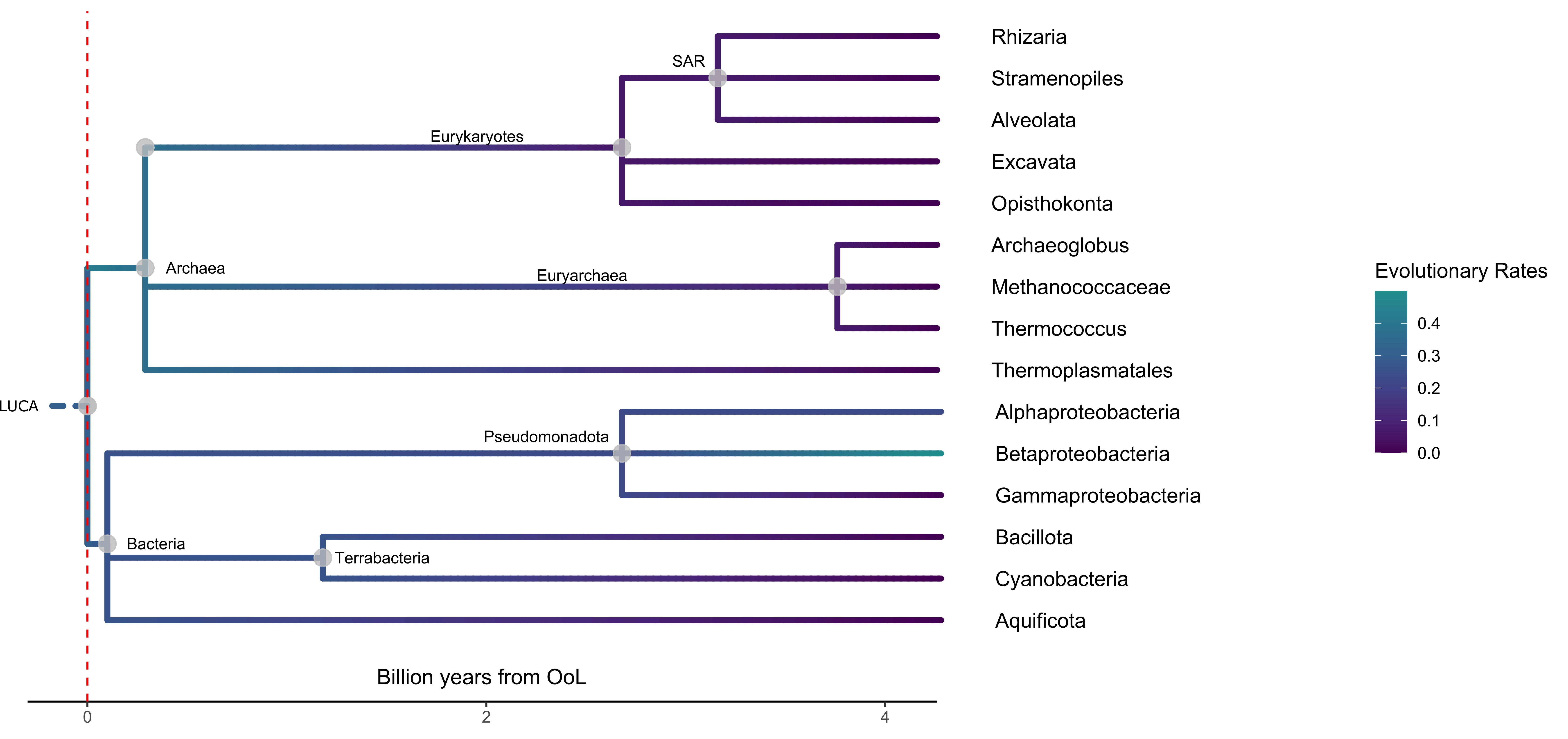}
         \includegraphics[width=17.8cm]{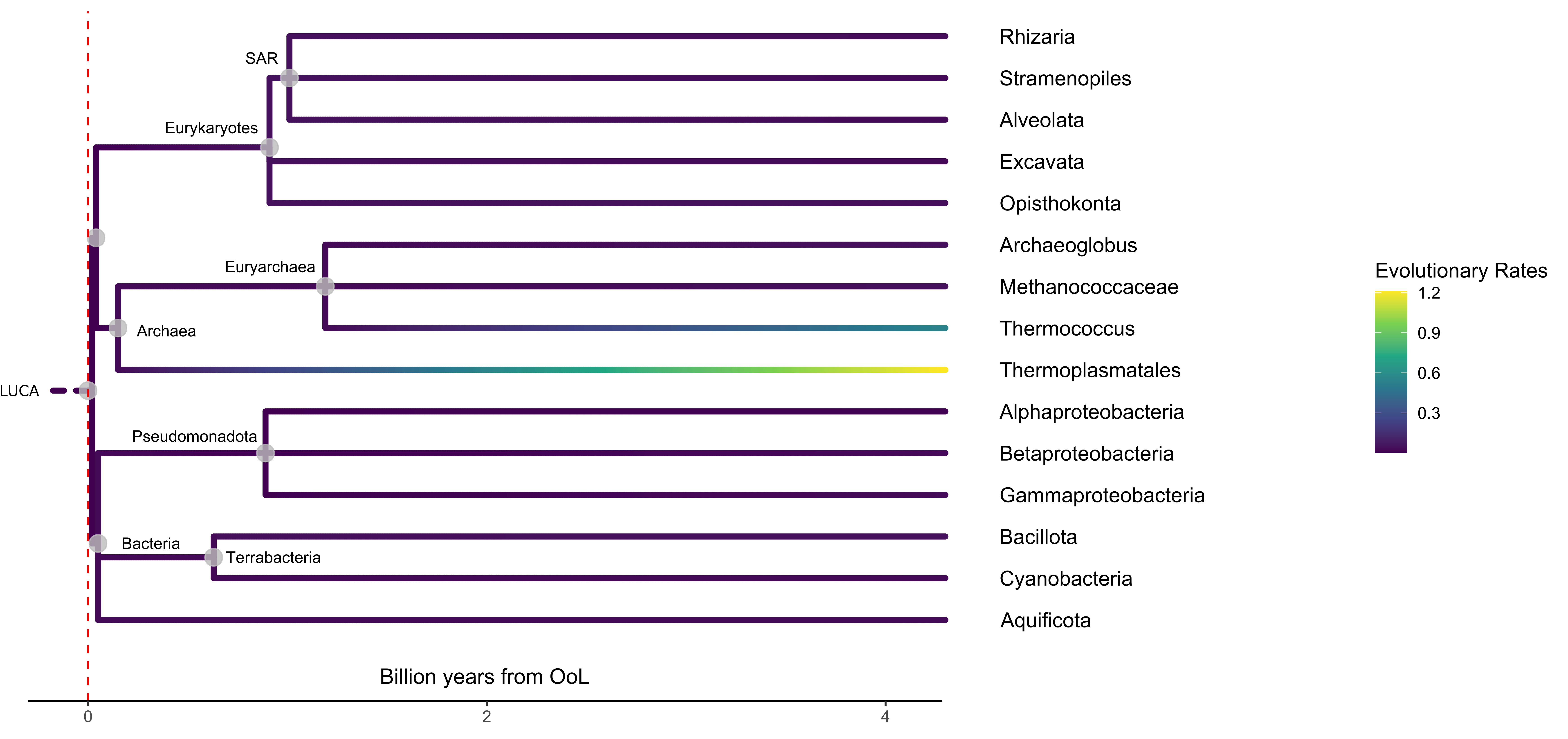}
        \caption{Time-calibrated phylogenetic trees with calculated evolutionary rates at Earth's median temperature (top) and at +10 K increase relative to Earth (bottom) with the colour indicating the evolutionary rates at the nodes.}
        \label{fig:treehot}
\end{figure*}

The dependence of the evolutionary rate on temperature has important consequences for the origination times of major unicellular groups across the tree of life. On Earth, the earliest domains to originate are the Bacteria and Archaea, within 0.31 Gyr after OoL, followed much later by the Eukaryotes at 2.7 Gyr. As shown in Figure \ref{fig:treehot}a, the different groups within these domains originated over a wide range of times, with the Terrabacteria originating at $\sim$1.1 Gyr while all the other groups originating beyond 3.2 Gyr. Overall, unicellular life in Earth's oceans remained relatively simple for nearly 2 Gyr after OoL, until the emergence of Eukaryotes thereafter marking the onset of complex life. Figure~\ref{fig:treehot}a also shows that the evolutionary rate across the different groups remained relatively uniform between $\sim$0.1-0.4 $\times$ $10^{8}$ mutations/nucleotide/sec, for our canonical case using an Earth-like temperature variation over time. We note that for a given temperature the evolutionary rates throughout the major unicellular groups are primarily driven by the body-size (Fig. \ref{fig:treehot})b, with larger species having lower metabolic rates and hence slower evolution. 

\begin{figure}
        \centering
          \includegraphics[width=0.48\textwidth]{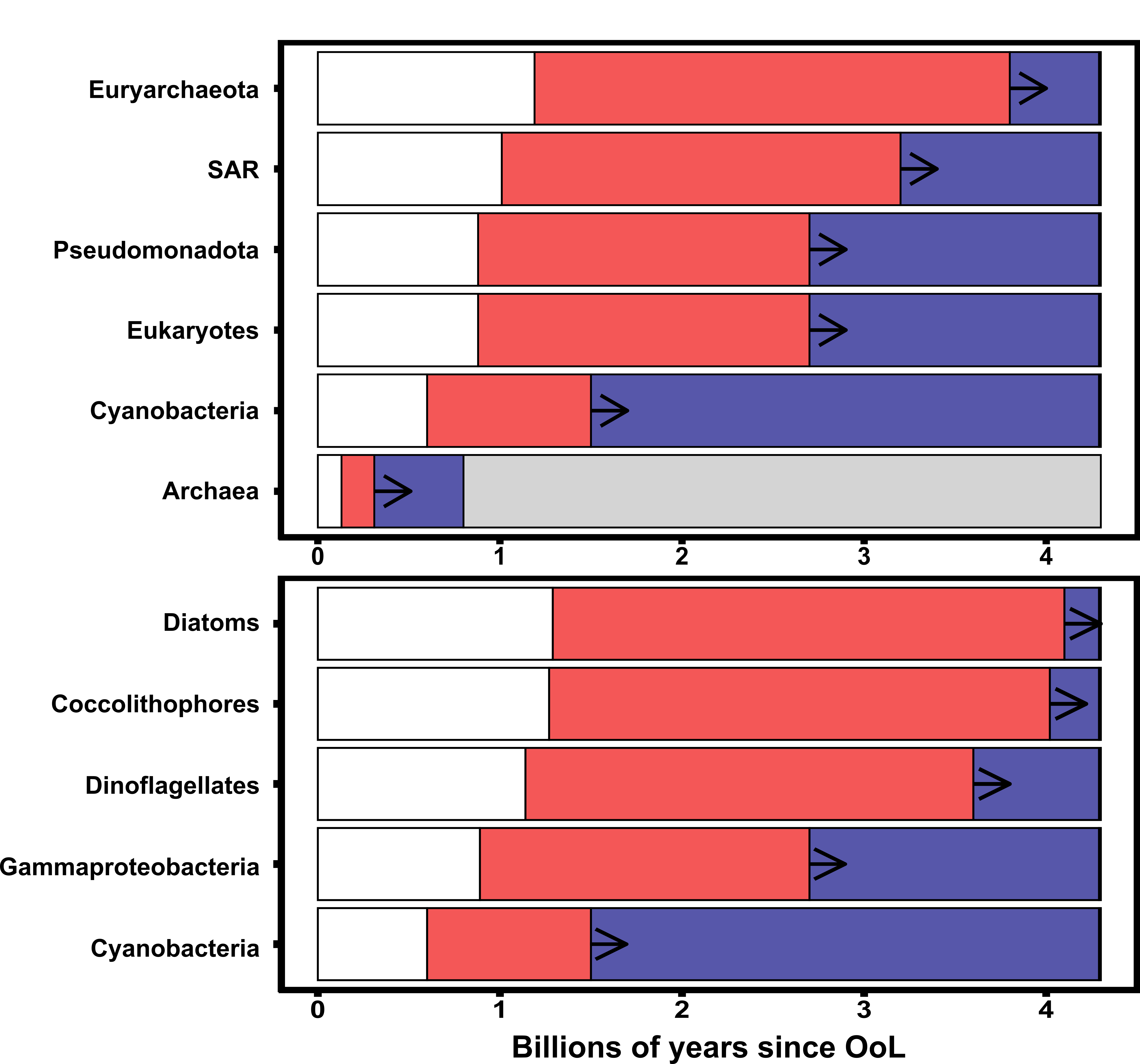}     
           \caption{Effect of temperature on origination times of major clades given in Fig. \ref{fig:treehot} (top) and of the key phytoplankton groups (bottom). Origination time on Earth of each group is marked with forward arrow. Red indicates increased temperature by +10 K and blue indicates decreased temperature by -10 K.}
         \label{fig:barmajor}
\end{figure}

A marginal increase in the median ocean surface temperature leads to significant changes in the origination times of different groups. Considering a median temperature of 10 K above that of Earth, the origination times of all the groups in the phylogeny happen significantly earlier, as shown in Fig. \ref{fig:treehot}b. In this case, all the major groups across all three domains now originate before 1.19 Gyr from OoL, with the Terrabacteria originating within 0.6 Gyr. The groups with the longest origination times in the canonical Earth-like case are affected more strongly than the others, because the effect of temperature change is cumulative. For example, the different Eukaryotic groups now originate within 1 Gyr compared to 3 Gyr in the canonical case, i.e. a 2 Gyr reduction in origination times. 
The changes in the origination times for the major groups for temperature differentials of $\pm 10 K$ can be seen in Fig. \ref{fig:barmajor}. The earlier origination times for all the groups with a 10 K increase in temperature is evident as discussed above. On the contrary, a similarly marginal decrease in the ocean temperature significantly delays the evolutionary rates and origination times for most of the groups. For a 10 K decrease the earliest species, Archaea, originates as late as 0.7 Gyr from OoL, compared to 0.2 Gyr for Earth, and most of the other groups originate later than 4 Gyr from OoL. 

Overall, we find that for Hycean worlds with temperatures marginally higher than ocean temperatures on Earth all the major groups in the phylogeny will have originated within 1.19 Gyr of OoL. These shorter origination times have important consequences for the prospects for biological evolution on such planets. Given the large planetary diversity in exoplanetary systems, and most with higher temperatures than on Earth, these findings indicate that at least microbial life could have evolved significantly earlier on such exoplanets than on Earth assuming such life formed in the first place. It also follows that for temperatures even higher than those considered here, i.e. $\Delta$T$>$ 10 K, the evolutionary rates could be even faster, until the temperatures become high enough to preclude survivability; on Earth most life is unexpected to survive for steady-state ocean temperatures above 40$^{o}$C \citep{sunday2012thermal}, albeit extremophiles are known to survive temperatures up to 120$^{o}$C \citep{dalmaso2015marine}. 

\subsection{Effect on Key Phytoplankton Groups}
We now focus on the evolutionary potential of several key phytoplankton groups, which are known to be abundant in the Earth's oceans and are key producers of biosignature gases in the Earth's atmosphere \citep{field1998primary}. Such species are of particular interest for Hycean worlds whereby phytoplankton in the oceans could produce molecules such as dimethyl sulfide that could be observable in their atmospheres. As discussed above, we consider five major groups of phytoplankton which originated throughout Earth’s history:  Cyanobacteria, Gammaproterobacteria (within the group Pseudomonadota), Dinoflagellates, Coccolithophores and Diatoms; these latter three belonging to the Eukaryotic group. Within the Bacteria domain, Cyanobacteria are the earliest group, originating 1.5 Gyr after OoL, followed by some Gammaproteobacteria at 2.4 Gyr. Within the Archaea domain, the three Eukaryotic phytoplankton groups originate relatively late with Dinoflagellates being the first at 3.6 Gyr, then Coccolithophores at 4.02 Gyr and Diatoms at 4.10 Gyr from OoL. For each of these groups we investigate how their evolutionary rates and origination times depend on the surface temperature. 

We find that the evolutionary behaviour of the phytoplankton groups follow a similar trend to other unicellular species considered above, i.e. that origination times are shorter with increased temperature. However, the key DMS phytoplankton groups are of consistently recent origin compared to other unicellular species in the Earth's history. For example, three of the five groups considered here, with the exception of Cyanobacteria and Gammaproteobacteria, originated relatively recently, 3.9 Gyr after OoL \citep{kumar2022timetree}. In particular, these later three groups are also the key DMS-producing phytoplankton in the present day oceans \citep{hopkins2023biogeochemistry,yoch2002dimethylsulfoniopropionate}. We find that an increase in the surface temperature by 10 K results in all the phytoplankton groups originating within 1.3 Gyr of OoL, as shown in  Fig. \ref{fig:barmajor}. In this case, Cyanobacteria have an origination time of only 0.25 Gyr post OoL. The Gammaproteobacteria follow shortly afterwards at 0.38 Gyr, and then the Eukaryotic phytoplankton groups all with origination times within 1.29 Gyr post OoL. 

On the contrary, a decrease in the surface temperature results in significantly slower evolutionary rate. With a marginal decrease in temperature by 10 K none of the key phytoplankton groups would originate before 4.3 Gyr from OoL, i.e. by the present age on Earth. On Earth, DMS production would have started with cyanobacteria at 1.5 Gyr, but it is not until 3.9 Gyr that significant DMS is produced by Dinoflagellates, followed by Coccolithophores then Diatoms at 4.10 Gyr.  However, on a planet with an increased surface temperature by 10 K relative to Earth, DMS could be started to be produced with cyanobacteria at 0.25 Gyr, with the major DMS producers by 1.28 Gyr.  Therefore the potential of a Hycean biosphere to produce large, detectable, amounts of DMS is present relatively early on in the planetary history for only a modest increase (+10 K) in temperature relative to Earth. 

 When the temperature change analyses were re-run using the Moody et al. 2024 dates, all key clades had evolved by 1.29 Gyr post OoL versus 1.28 Gyr post OoL with with the time tree data and the key phytoplankton clade had evolved by 1.16 Gyr post Ool versus 1.19 Gyr post OoL.  As such, our Time-Tree data model represented a slightly more conservative estimate of these origination times. 

\section{Summary and Discussion}
In this study we investigated the prospects for biological evolution of unicellular life in oceanic environments of Hycean worlds at different median surface temperatures compared to Earth. We find that relatively marginal changes in ocean surface temperatures compared to Earth's surface temperature over planetary timescales can lead to significant changes in the evolutionary rates and origination times of important species. For example, a 10 K increase relative to Earth leads to evolutionary rates which are over twice as fast, while a decrease of 10 K halves them.  Faster evolutionary rates lead to faster speciation times meaning that the time it takes for new groups of species to form reduces.  This increased rate has a significant impact on the origination times of unicellular groups such that for an increase of 10 K of surface temperature, all of the major groups will have originated by 1.19 Gyr post OoL and all the key phytoplankton groups by 1.28 Gyr. In contrast, a decrease of 10 K of median surface temperature severely limits the origination rates, such that by 4 Gyr post OoL only Bacteria and Archaea will have evolved, but not oxygenic photosynthesis or Eukaryotes. 

\subsection{Implications}
A central finding of our study is that a large range of evolutionary rates and origination times are possible within our model framework for unicellular life in oceanic environments for a relatively marginal range in environmental conditions. First, given the wide range of possible atmospheric conditions in Hycean worlds, an equally wide diversity in microbial life could be expected. In particular, the origination of new clades in warm Hycean worlds can happen significantly faster than on Earth. This faster origination is of particular relevance for currently known candidate hycean worlds, all of which are expected to host ocean temperatures significantly warmer than the Earth if at all they are habitable 
\citep{madhusudhan_habitability_2021}. Secondly, warm Hycean worlds that are significantly younger than Earth could also provide the conditions for originating and sustaining major unicellular groups. Within our model considerations, a marginal (10 K) increase in mean surface temperature relative to Earth can lead to origination of major unicellular groups as early as $\sim$1 Gyr from OoL. Therefore, candidate Hycean worlds orbiting stars significantly younger than the sun, such as K2-18 b at 2.4 billion years old \citep{guinan2019k2},  could also provide important targets for biosignature searches. 

Our results also indicate an optimal range of ocean surface conditions that may lead to significant atmospheric biosignatures. Planets with relatively colder surface temperatures can significantly inhibit evolutionary transitions and delay the origins of important microbial groups. In particular we find that a lower median surface temperature that is 10 K below that of Earth value could cause a delay in the origination times of key phytoplankton groups by several Gyr. This delay in turn could affect their observable biomarkers such as DMS, which is known to be produced predominantly by Eukaryotic phytoplankton in Earth's oceans which originated relatively late compared to other microbial species. Therefore, ocean worlds that are significantly cooler than Earth may be expected to host simpler microbial life (in terms of phyto-autotrophic systems) than Earth's oceans and so may show weaker biosignatures, unless they orbit significantly older stars than the Sun. Conversely, habitable ocean worlds with warmer surface temperatures are more likely to show stronger atmospheric biosignatures due to microbial life if present.  

\subsection{Limitations}
We have performed a conservative study on the evolutionary rate of well-studied unicellular organisms in the Earth’s oceanic environment over the history of life on Earth, with a single species representing each broad taxonomic group. There are some caveats to our study. First, the groups themselves are evolutionarily stable even though the species within them could speciate and go extinct, and so our results are limited to the behaviour of groups, rather than individual species. Secondly, the implications for our model should not be extended to multicellular complex organisms because we parameterised our model using empirically derived data for unicellular organisms, and so the dynamics of multicellular life are not necessarily represented by our model. Thirdly, the temperature range we considered was the normal thermal tolerance for life, and while there is evidence that the MTE holds for extremophiles \citep{mcclain2012energetics}, the variation/differences/constraints of the MTE relationships are not as well studied. Therefore, caution should be exercised when extending the results to temperatures above 313 K ($40^\circ$ C). Finally, we have used temperature as the key parameter influencing the origination times for the species we consider in this work. However, other factors such as the availability of bioessential chemicals may also need to be considered in future studies. 

\subsection{Future work}
Within the framework of the present study we find the median ocean surface temperature and mass of the organism to be key determinants of the origination time for the predominant unicellular species considered. Our choices for the temperature and physical conditions considered were motivated by those of the Earth's oceans. Therefore, our findings may only represent nominal estimates considering that a much broader environmental diversity is expected for extraterrestrial habitable planets. Future work in this direction could explore a range of other conditions, including the effect of gravity, pressure, larger temperature variations and other environmental factors. Similarly, future studies could also investigate the implications for more complex life, beyond the unicellular life considered in this initial study, including multicellular/animal life in similar conditions as well as simpler life in more extreme conditions, i.e. extremophiles. Finally, while our focus in the present work has been exoplanetary Hycean worlds the results may also be extended to other habitable environments. Such habitats could include oceans on habitable rocky exoplanets,  sub-surface oceans on icy moons in the solar system, as well as other ocean worlds with varied atmospheric compositions. 

In conclusion, while our model makes some broad  assumptions, the overarching patterns that emerge show that within our framework, for unicellular organisms, changes in median surface temperatures can have large impacts on the evolutionary rates and origination times of major groups. These changes mean that with different surface temperatures, the biosphere could be relatively complex at a young age for warmer planets, or relatively simple at an old age for cooler planets. Such biospheres with varied levels of complexity can impact the detectability of life on them, such that warmer planets have the potential to show strong atmospheric biosignatures.
\section*{Acknowledgements}
EGM acknowledges support from a NERC IRF (2019-2024): NE/S014756/1. NM acknowledges support from UKRI Frontier Grant: EP/X025179/1. This research has made use of the NASA Exoplanet Archive, which is operated by the California Institute of Technology, under contract with the National Aeronautics and Space Administration under the Exoplanet Exploration Program.

Author Contributions: EGM and NM conceived and planned the project. EGM performed the numerical calculations. EGM and NM discussed and synthesized the results and wrote the manuscript.

\section*{Data Availability}
This is a theoretical work and no new data was generated as a  result. 
 



\bibliographystyle{mnras}
\bibliography{references} 




\appendix

\section{Model Assumptions}
\label{appendix:assumptions}
Robustly modelling the evolutionary history of life on a planet is a complex problem, and that is not fully understood on Earth. In order to understand the nature of life that may exist on other planets, we would need to fully understand the nature of both the origins of life and biological evolution over planetary timescales.  Yet, these are still very active areas of research, and fundamental questions, such as the extent to which evolution is deterministic, remain unanswered on Earth.  As such, our approach involves working with neutral empirical models which capture macroevolutionary patterns well on Earth while minimising assumptions and parametrisation \citep[e.g.][]{budd_dynamics_2020,mitchell_importance_2019}. Therefore, we consider the simplest set of assumptions to provide a baseline of what the evolution of unicellular life on Hycean planets could look like. We consider only temperature as a free parameter and only the key unicellular groups that are well known from the evolutionary history of Earth. 

The assumptions we need to make in order to make our theoretical predictions can be grouped into four categories: 1) assumption that life has originated, 2) conceptual assumptions about how evolution works, 3) input parameters for the MTE equations, and 4) differences that could occur on different planets.

First, because the conditions for life to have originated on Earth are very much debated \citep{yvcas1955note,rich1962problems,sasselov2020origin}, we have made the assumption that it can and has started within our model.  We also assume that once life has started, the environmental conditions are such that life can be maintained over the planetary timescales. The biosphere today is a complex set of interactions and feedbacks such that perturbations to the biosphere are adjusted, helping to maintain long-term stability. Prior to the evolution of complex, multicellular life, i.e. animals, on Earth 600 million years ago, such large-scale feedbacks were likely more limited, yet the biosphere was still stable. Life was maintained for billions of years before animals developed the capacity to adapt and change the environment for their own benefit.  As such, assuming a stable biosphere for unicellular life, such as the one we have had on Earth, is a reasonable assumption.  

Secondly, the assumptions with the most uncertainty are those that we need to make about how biological evolution works. It is currently unknown the extent to which, once life has started, the Earth patterns of evolution are deterministic or subject to some probabilistic pattern. There could also be certain bottlenecks that require a specific set of conditions in order for the next major evolutionary transitions to occur. Our model assumes the independence of different groups (i.e. no biotic or abiotic interactions). However, we know that key events such as the evolution of oxygenic photosynthesis led to diversification in bacteria \citep{davin2023evolutionary} likely through abiotic selection pressures and/or the opening of niches and through biotic interactions. On the other hand, the $\sim$1.2 billion years between the rise of oxygen and the origins of Eukaryotes \citep{olejarz2021great} suggests that it is unlikely that oxygen directly drove the evolution of complex (i.e. Eukaryotic) life. Because oxygen is unlikely to be readily available in the H$_2$-rich atmospheres of Hycean worlds it is unclear whether bacterial life would follow a similar evolutionary trajectory as on Earth but with different selection pressures or a different, possibly less diverse, trajectory altogether. Other key events also likely contributed to diversification, such as the origin of metazoan zooplankton resulting in predation pressure and leading to phytoplankton diversification \citep{butterfield1997plankton}.  If one of these key major evolutionary transitions didn’t happen, it is not clear what the impact is on the other groups. These are all major questions in evolutionary biology, and so to minimise assumptions of what key drivers are, and the likelihood that major evolutionary transitions occur, we take the simplest neutral model for this study, namely we only consider unicellular life and that evolution will proceed similarly to Earth’s as a starting point for such Exo-evolutionary studies.

Thirdly, we have to parameterise our MTE equations, which we know have variations within different taxonomic groups.  We have limited the impact of such parametrisation by focusing on unicellular organisms only, and discussing relative evolutionary rates, such that within a taxonomic group that has the same parameter set (e.g. within unicellular organisms), the absolute value will not change the results. This approach means that inferences for multicellular life should not be made based on our models. Our analyses also rely on observed (and calculated) values of the Earth’s temperature, the origination times of groups, the body sizes of the model organisms for each groups and the topography for the phylogeny used. 

Origination times for different groups vary depending on the methodology used, which is why we used a model averaging approach \citep{kumar2022timetree} which enabled us to cover many different groups within the same approach. We also performed some sensitivity analyses using a more recent phylogeny \citep{moody2024nature}, to assess how changing origination times impacted results.  Broadly speaking the more recent the origination time, the smaller the confidence interval (e.g. \citet{moody2024nature}), while the earliest originating groups tend to have much larger confidence intervals, i.e. are not as precisely known.  This pattern works in our favour because the longer a group takes to evolve, the greater the impact of a change of median temperature, but the smaller the origination time uncertainty, so that the groups with the largest uncertainty, i.e the oldest ones will only have smaller changes to their origination times with different temperatures. 

The topography of phylogenies can be highly variable with different methods and data e.g. \citep{williams2020phylogenomics, moody2024nature}, so we have worked with broad taxonomic groups which provides robustness to the variations in phylogeny topology that occur at a finer taxonomic scale. Sensitivity analyses on having three versus two domains showed only limited impact on the origination times of the groups, e.g. Eukaryotes changed origination time by 0.01 Gyr. 

Organism body sizes vary, but because the metabolic rate scales with mass$^{-\frac{1}{4}}$ the impact of small uncertainties on our model are limited.  However, because of the sensitivity of our model to temperature, the biggest source of practical (rather than conceptual) uncertainty comes from the resolution of the median Earth temperature.  We used the median values from \citet{krissansen2018constraining}, with their 95\% confidence intervals of $\pm$20 K at 4 billion years ago but decreasing through time to a few degrees uncertainty for the last 500 million years. Considering the full range of uncertainties at the beginning may have a significant impact on the results, which currently consider a maximum variation of $\pm$15 K relative to Earth at any given time. However, using values at the limits of this confidence interval would also involve modelling outside the normal thermal tolerance of life, so would introduce more assumptions and uncertainty. As such, using the median surface temperature is a conservative approach.

Finally, we have assumed that the production of energy by unicellular life on Hycean worlds would operate similarly to life on Earth, namely carbon-based, nucleotide based systems. Our model assumed such Earth-like life as the most conservative since other forms of life, such as silicon-based life, while possible is more challenging to occur than carbon-based life \citep{petkowski2020potential}.  Yet different planets will have different gravities, different stellar flux and potentially different types of dominant photosynthesis. Predicting how different gravity may change evolutionary rates and origination times is complex because the best proxy we have for gravity is the high-pressures that we get in the deep-sea.  While deep-sea organisms can exhibit higher speciation rates \citep[e.g. ][]{martinez2021deep}, these appear primarily driven by biotic not abiotic factors because the internal pressure of organisms is not dramatically reduced (~ 1-2 \% for 4km deep). Therefore, the impact of increased pressure is likely much smaller than the biotic factors that likely drive the increased evolutionary rates, so is a more complex dynamics. Calculations of how different stellar spectra change input power have demonstrated that predicted variation is within an order of magnitude for cool stars \citep{duffy2023photosynthesis}, suggesting limited influence compared to the impact of temperature, as considered in this work. 

\bsp	
\label{lastpage}
\end{document}